# Can corrections to gravity at galactic distances be decisive to the problem of dark matter and dark energy?

**T.F. Kamalov[1,3,4], O.A. Volkova[1], M.H. Khamis Hassan[1,2], N.V. Evdokimov[1] and Yu. T. Kamalov[1]**

*[1] Department of Theoretical Physics, Moscow Region State University, Moscow Region, Mytishchi, 141014, Russian Federation*
*[2] Department of Physics, Faculty of Science, Assiut University, Assiut, 71511, Egypt*
*[3] Department of Theoretical Physics, Moscow Institute of Physics and Technology, Moscow Region, Dolgoprudny, 141701, Russian Federation*
*[4] PushGENI Branch RosBiotex University, Prospect Nauki, 3, Puschino, Moscow Region, 142290, Russian Federation*

E-mail: kamalov@gmail.com, olka.volkova96@yandex.ru, m.khamis@yandex.ru, timkamalov@gmail.com, nv.evdokimov@gmail.com

**Abstract -** Are Dark Matter the result of uncalculated addition derivatives? The need to introduce dark matter dark becomes unnecessary if we consider that, the phenomenon of dark matter is a result of not computing the additional derivatives of the equation of motion. For this purpose, we use higher derivatives in the form of non-local variables, known as the Ostrogradsky formalism. As a mathematician, Ostrogradsky considered the dependence of the Lagrange function on acceleration and its higher derivatives with respect to time. This is the case that fully correspond with the real frame of reference, and that can be both inertial and non-inertial frames. The problem of dark matter and dark energy presented starting from basic observations to explain the different results in theory and experiment. The study of galactic motion, especially the rotation curves, showed that a large amount of dark matter can be found mainly in galactic halos. The search for dark matter and dark energy has not confirmed with the experimental discovery of it, so we use Ostrogradsky formalities to explain the effects described above, so that the need to introduce dark matter and dark energy disappears. We hope that this description can lead to a Steady State Model (SSM) of the Universe.

## 1. INTRODUCTION

Gravity correction at galactic distances can decisively solve the problem of dark energy. Dark energy is a theoretical concept that is used to explain the observed accelerated expansion of the universe, and it is a much broader issue than just explaining the motion of galaxies.

While gravity is the dominant force at galactic distances, and it does play a role in shaping the motion of galaxies, it can account for the observed acceleration of the universe's expansion. Dark energy is believed to be is a quantity that opposes the attractive force of gravity and is causing the expansion of the universe to accelerate.

There have been various proposals to modify the laws of gravity to account for the observed acceleration of the universe's expansion, such as modifying general relativity or introducing new fields that interact with matter [1-5]. However, these proposals are still in the realm of theoretical physics and have not yet been confirmed by observations [6-9].

Therefore, gravity correction at galactic distances may help us better understand the dynamics of galaxies, it can decisively solve the problem of dark energy, which is a much broader and complex issue. Discussions related to this issue can be found in [10-16].

In 1933, he studied the redshifts of various galaxy clusters, as published by Edwin Hubble and Milton Humason in 1931, and noticed a large scatter in the apparent velocities of eight galaxies within the Coma Cluster, with differences that exceeded 2000 km/s. The fact that Coma exhibited a large velocity dispersion with respect to other clusters had already been noticed by Hubble and Humason, but Zwicky went a step further, applying the virial theorem to the cluster in order to estimate its mass.

This was not the first time that the virial theorem, borrowed from thermodynamics, was apply to astronomy. As to the best of our knowledge, Zwicky was the first to use the virial theorem to determine the mass of a galaxy cluster.

Zwicky started by estimating the total mass of Coma to be the product of the number of observed galaxies, 800, and the average mass of a galaxy, which he took to be 109 solar masses, as suggested by Hubble. He then adopted an estimate for the physical size of the system, which he took to be around 106 light-years, in order to determine the potential energy of the system. From there, he calculated the average kinetic energy and finally a velocity dispersion. He found that 800 galaxies of 109 solar masses in a sphere of 106 light-years should exhibit a velocity dispersion of 80 km/s. In contrast, the observed average velocity dispersion along the line-of-sight was approximately 1000 km/s. From this comparison, he concluded: If this would be confirmed, we would get the surprising result that dark matter is present in much greater amount than luminous matter [6]. The study of the rotational motion of spiral galaxies carried out by Vera Rubin [17] in the early 1970s of the last century showed that the rotational velocity of the outer ends of the Andromeda Galaxy, which is the closest to us, is almost constant for all its outer parts and does not depend on the distance from the center. This is what puzzles astrophysicists. As it contradicts expectations that the speed of movement of the outer ends of the galaxy must depend on its distance from the center of the galaxy.

## 2. QUANTUM CORRECTION TO NEWTON'S LAW

From Ostrogradsky formalism using a Lagrange function is

$$L = L(q, \dot{q}, \ddot{q}, \dots, q^{(n)}, \dots), \tag{1}$$

but not

$$L = L(q, \dot{q}). \tag{2}$$

The Euler–Lagrange equation with high-order addition variables follows from the least-action principle:

$$\delta S = \delta \int \mathrm{L}\left(\mathrm{q}, \dot{\mathrm{q}}, \ddot{\mathrm{q}}, \dots, q^{(n)}\right) \mathrm{dt} = \int \sum_{n=0}^{N} (-1)^n \frac{\partial^n}{\partial t^n} \left(\frac{\partial L}{\partial q^{(n)}}\right) \delta q^{(n)} dt = 0 \tag{3}$$

This equation can rewritten in the form of a corrected Newton's second law of motion in non-inertial reference frames [18]:

$$F - ma + f_0 = 0. \tag{4}$$

Here,

$$f_0 = mw = w(t) + \dot{w}(t)\tau + \sum_{k=2}^{n}(-1)^k \frac{1}{k!}\tau^k w^{(n)}(t) \tag{5}$$

is a random inertial force [18] that can be represented by Taylor expansion with high-order derivatives coordinates on time

$$F - ma + \tau m\dot{a} - \frac{1}{2}\tau^2 ma^{(2)} + \ldots + \frac{1}{n!}(-1)^n\tau^n ma^{(n)} + \ldots = 0 \tag{6}$$

in inertial reference frame w = 0.

In Newtonian case

$$F = G\frac{mM}{r^2} \tag{7}$$

It follows from the equivalence principle of gravity and inertia that Newton's second law extended to random non-inertial frames of reference should also add additional variables to the law of gravitational interaction. On the other hand, it follows from the ergodic hypothesis that the time averages are equal to their average statistical values $r$ [18]. Therefore

$$ma - \tau m\dot{a} + \frac{1}{2}\tau^2 ma^{(2)} - \ldots + \frac{1}{n!}(-1)^n\tau^n ma^{(n)} + \ldots = m\frac{GM}{r^2}\left(1 - \frac{\lambda}{r} + \frac{\lambda^2}{r^2} - \cdots\right) = m\frac{GM}{r^2}e^{\left(-\frac{\lambda}{r}\right)}, \tag{8}$$

here $\lambda$ is measure of $r$.

## 3. DARK METRIC FOR DARK MATTER and DARK ENERGY

It follows that the phase space of coordinates and high-order derivatives gives the corrected Newton's formula for gravitational potential [19]

$$\varphi = \varphi_0 e^{-\frac{\lambda}{r}} \tag{9}$$

where $\varphi_0 = \frac{GM_g}{r_g}$, potential; $G$, gravitational constant and $M$, mass.

In our case

$$G\frac{mM_g}{r_g^2}e^{-\frac{\lambda}{r}} \approx \frac{mv^2}{r_g}, \tag{10}$$

then

$$v \approx \sqrt{\frac{GM_g}{r_g}}e^{-\frac{\lambda}{r}} \tag{11}$$

because the correction coefficient $e^{-\frac{\lambda}{r}}$ for gravity, $r_g$ and $M_g$– radius of Galactic rotation and mass of Galactic.

On the one hand, force F is expressed using infinite Taylor expansion. On the other hand, gravitational force $F_g$ can also represented as a series, as follows from the principle of equivalence. If this series is replaced by an exponential, then we can write metric

$$ds^2 = e^{-r_0/r}dt^2 - e^{r_0/r}dr^2 - r^2 d\theta^2 - r^2 sin^2\theta d\phi^2$$

which we call the dark metric [18], where $r_0 = 2GM/c^2$.

The dark metric is the asymptotic of the Schwarzschild metric for $r_0 < r$. The definition of dark metrics for matter and energy presented to replace the standard notions of dark matter and dark energy.

The dark metric can also obtain from the standard metric:

$$ds^2 = B(r)dt^2 - A(r)dr^2 - r^2 d\theta^2 - r^2 sin^2\theta d\phi^2$$

Conditions $A(r)B(r) = 1$ and $limA(r) = B(r) = 1$ for $r \to \infty$ must be satisfied for the standard metric. The dark metric also satisfies to these conditions. Gravitational forces are presented as a series with changing signs *(Fig. 1, Fig. 2)*.

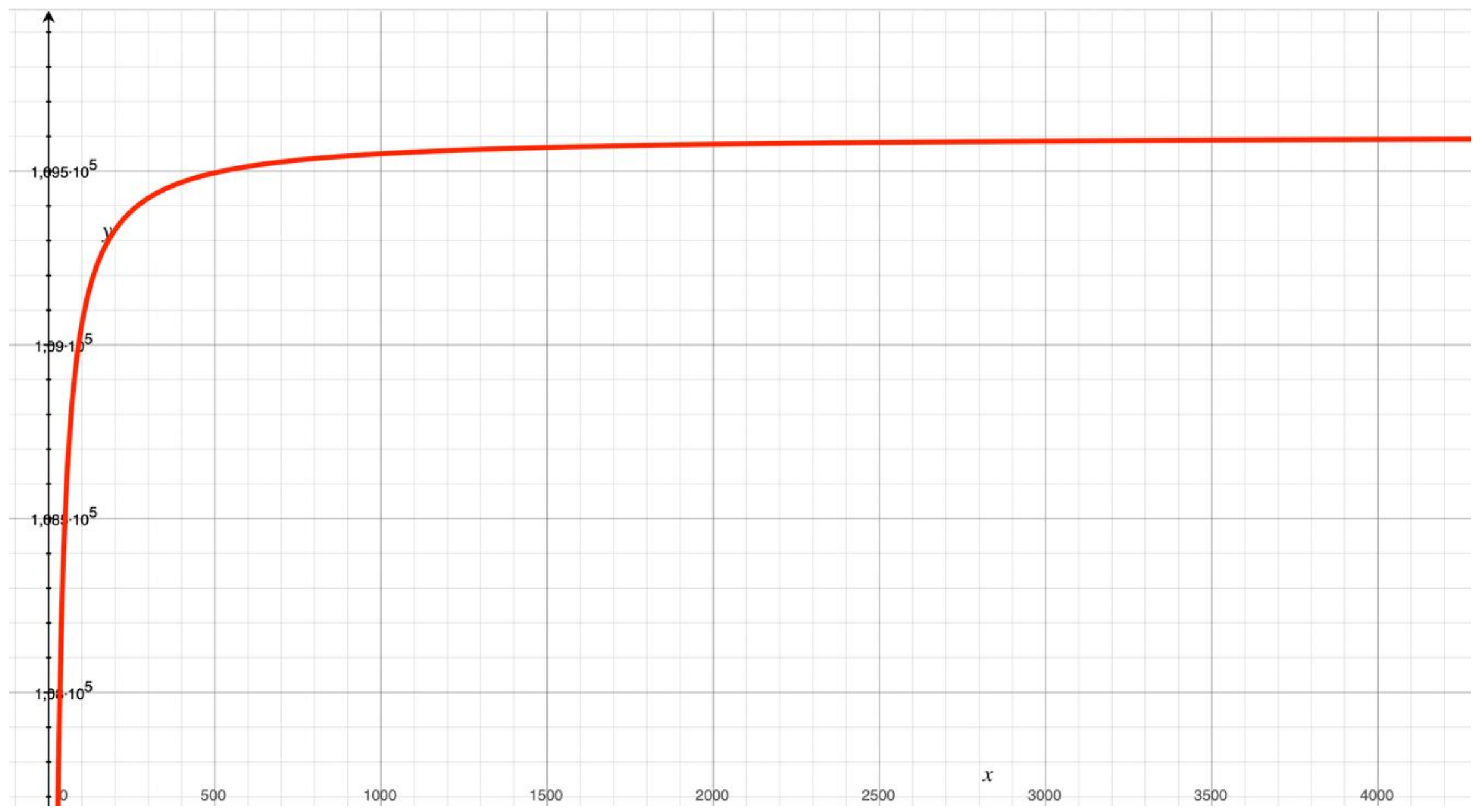

*Fig. 1. The curves velocity rotation* $v = \sqrt{\frac{GM_g}{r_g}}\, e^{-\lambda/2r} = \sqrt{\frac{6{,}674\cdot10^{-11}\cdot9\cdot10^{40}}{5\cdot10^{20}}}\cdot e^{-\frac{1}{2r}}$, *(m/s) of Milky Way Galaxy depends from radius of rotation* $r$ *(kpc).*

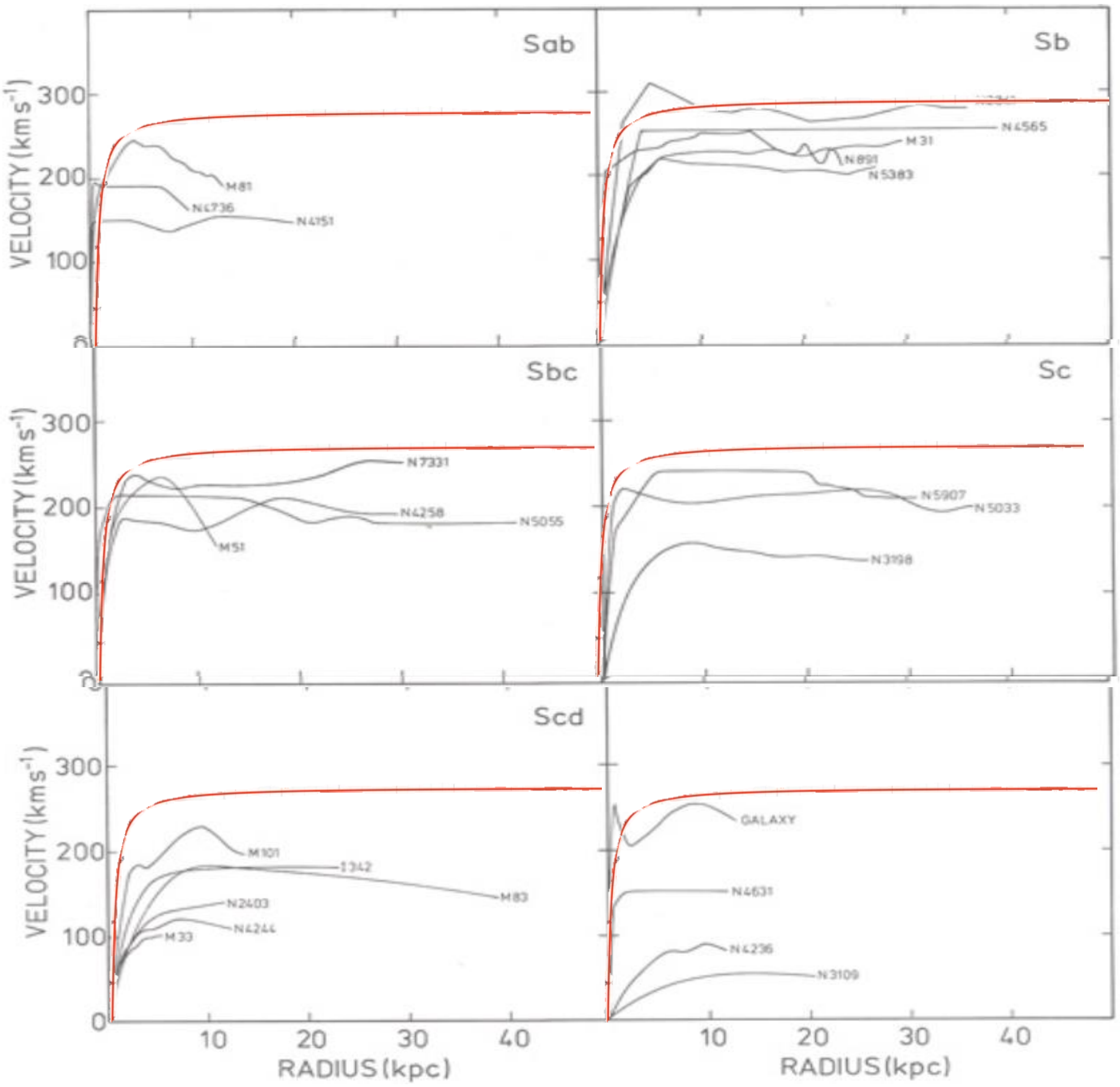

*Fig. 2. The rotation curves of the 25 galaxies published by Albert Bosma in 1978 [6] (red is our theoretical result).*

## 4. MODEL OF STEADY STATE UNIVERSE

Now let's set the task: "what should be the formula in order for the experimental data to coincide with the theoretical ones?" According to the Doppler effect, one can write

$$\omega = \omega_0\left(1 - \frac{Hr}{c}\right) \quad (12)$$

where $H$ is the Hubble constant, $\omega_0$ is the angular frequency with which the source emits waves, $c$ is the speed of light in vacuum, $v = Hr$ is the removal velocity of galaxies proportional to their distance.

The corresponding energy is

$$E = \hbar\omega_0\left(1 - \frac{Hr}{c}\right) \quad (13)$$

where $\hbar$ is Planck's constant.

Taking into account the correction to the gravitational interaction at galactic distances and for

replacing $\lambda$ to $r_1$ and $rr$ to $r_2$ , we obtain

$$E = E_0 e^{-\frac{t - r_1/c}{t - r_2/c}} \quad (14)$$

The energy and frequency are respectively equal

$$E \approx E_0\left(1 - \frac{t - r_1/c}{t - r_2/c}\right), \quad (15)$$

$$\omega \approx \omega_0\left(1 - \frac{t - r_1/c}{t - r_2/c}\right) \quad (16)$$

and from $r_1 = 900\ kpc$ corresponds to dark energy, $\omega_0 = \frac{2\pi c}{\lambda}$ (Fig. 3).

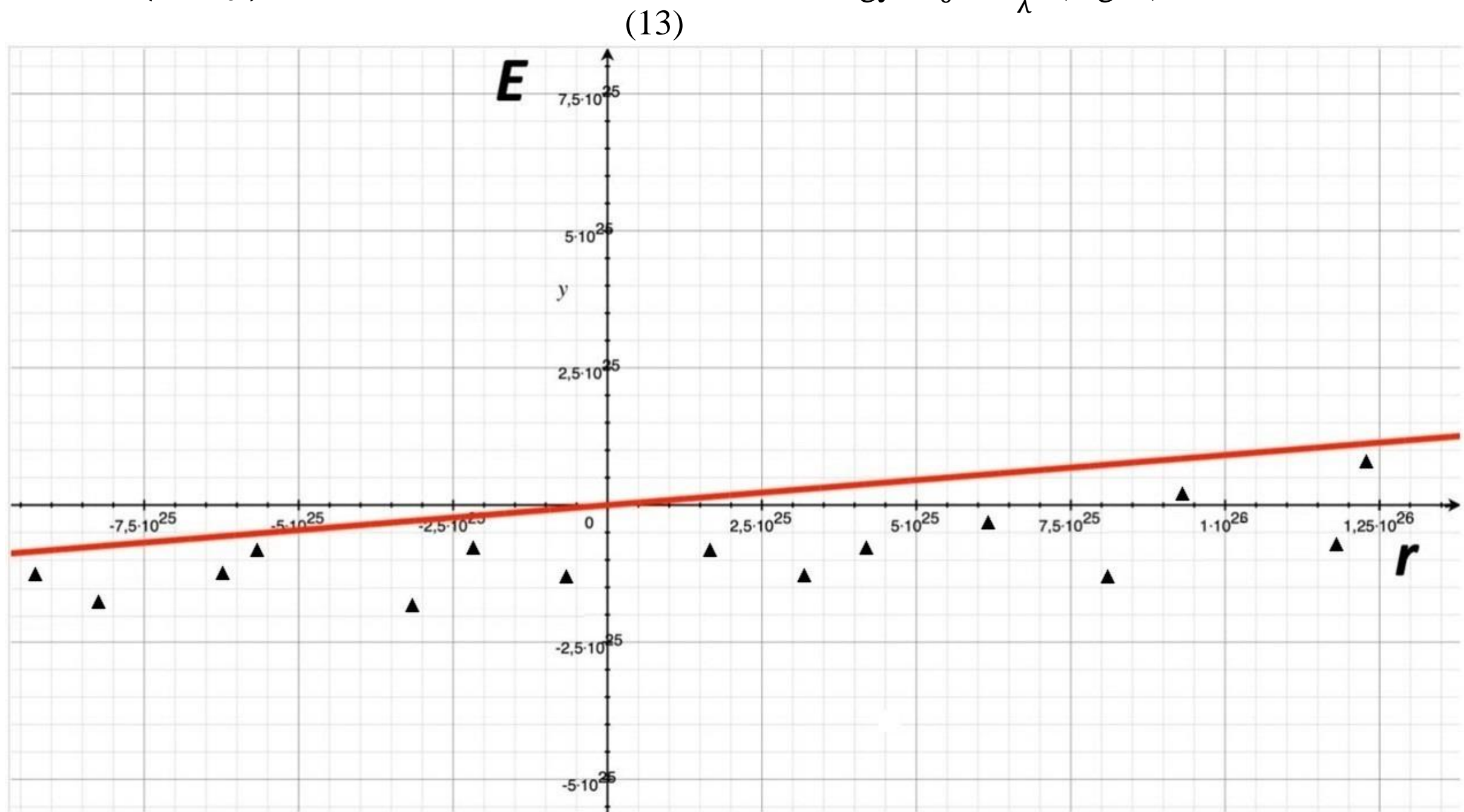

*Fig. 3. Graphs of the dependence of the energy and frequency with which the source emits waves on the distance from the source to the observer* $E = \hbar\omega_0\left(1 - \frac{Hr}{c}\right)$ *(the dotted line shows the description of the redshift using the Doppler effect) and* $\omega \approx \omega_0\left(1 - \frac{t - r_1/c}{t - r_2/c}\right)$*, where* $r_2 = -2{,}772 \cdot 10^{22}$ *(the graph obtained taking into account the correction we introduced is shown by the red line).*

## 5. CONCLUDION

In the general case, non-inertial dynamics can describe by high order differential equations. From the principle of equivalence, it follows that the gravitational force also has to be represent as a series. The corresponding metric called the dark metric. The dark metric describes gravitational interaction with additional terms that lead to the description of observable effects of dark matter and dark energy. This means that the correct calculation using the dark metric leads to an abandonment of notions of dark matter and dark energy. Therefore, there is no need to seek for something that does not exist. The proof of this statement is the good agreement between our theoretical corrections Newton Law and experimental data. We hope that the gravity correction at galactic distances can be decision the problem of Dark Matter and Dark Energy and follows to the Steady State Model (SSM) of Universe.